\documentclass[draftcls,10pt]{ismrm}
\usepackage{endfloat}
\usepackage{makeidx}  
\usepackage{amsmath}

\usepackage{graphicx}
\usepackage{array}
\usepackage{bm}
\usepackage{color}
\usepackage{amssymb}
\usepackage{amsmath,graphicx,subfigure}
\usepackage{widetext}

\def\W{{\cal W}}

\def\F{{\cal F}}
\makeatletter
\renewcommand\tagform@[1]{\maketag@@@{\ignorespaces#1\unskip\@@italiccorr}}
\makeatother

\def\bk{$\boldsymbol k$}
\def\bx{$\boldsymbol x$}
\def\snr{\text{SNR}}

\title{Exact Calculation of Noise Maps and g-Factor in GRAPPA using a \bk--space Analysis (Submitted to Magnetic Resonance in Medicine)}
\author{\large I\~{n}aki Rabanillo-Viloria, Santiago Aja-Fern\'andez, Carlos Alberola-L\'opez\\
\normalsize LPI, ETSI Telecomunicaci\'on, Universidad de Valladolid, Spain\\
\normalsize Email: irabvil@lpi.tel.uva.es\\
\large Diego Hernando\\
\normalsize Departments of Radiology and Medical Physics, University of Wisconsin, Madison, WI} 
\runningtitle{Rabanillo-Viloria et al.}{Noise Characterization in GRAPPA using a \bk--space Analysis} 
\wordcount{4966}
\begin{document}

\maketitle

\begin{abstract}

Noise characterization in MRI has multiple applications, including quality assurance and protocol optimization. It is particularly important in the presence of parallel imaging acceleration, where the noise distribution can contain severe spatial heterogeneities. If the parallel imaging reconstruction is a linear process, an exact noise analysis is possible by taking into account the correlations between all the samples involved. However, for k-space based techniques like GRAPPA, the exact analysis has been considered computationally prohibitive due to the very large size of the noise covariance matrices required to characterize the noise propagation from k-space to image-space. Previous methods avoid this computational burden by approximating the GRAPPA reconstruction as a pixel-wise linear operation performed in the image-space. However, these methods are not exact in the presence of non-uniform k-space undersampling (e.g.: containing a calibration region). For this reason, in this work we develop an exact characterization of the noise distribution for self-calibrated parallel imaging in the presence of arbitrary Cartesian undersampling patterns. By exploiting the symmetries and separability in the noise propagation process, the proposed method is computationally efficient and does not require large matrices. In this manuscript, we present the proposed noise characterization method and compare it to previous techniques using Monte-Carlo simulations as well as phantom acquisitions.
\end{abstract}


\section{Introduction}


Thermal noise is an unavoidable source of degradation in MRI signals. The principal source of noise is the subject itself, followed by electronic noise during the acquisition of the signal in the receiver chain~\cite{AjaNoiseBook}. Noise degrades the visual quality of the reconstructed images, and complicates further post--processing techniques, such as segmentation, registration, fMRI analysis or numerical estimation of parameters. Accurate characterization of noise statistics is essential for many different tasks such as quality assurance \cite{Krissian09,Aja13}, protocol optimization \cite{Thunberg07,Saritas11}, and tailoring of subsequent post--processing steps \cite{Rabanillo16,Ghugre05,Veraart11}.  


For practical purposes, noise in the \bk--space is usually assumed to be a zero-mean, spatially uncorrelated independent and identically distributed (IID) complex Gaussian process for each coil, with equal variance in both the real and imaginary parts. If the data is acquired by several receiving coils, the multi-coil noise can be characterized by its covariance matrix.    


For linear image reconstructions, the propagation of noise in \bk--space into the image--space (also known as \bx--space) can be described by matrix operations. In the case of fully-sampled acquisitions, the IID noise behavior is preserved when the data are transformed into the \bx--space, due to the orthogonality of the inverse Fourier Transform (iFFT). However, in the presence of multiple coils and parallel MRI (pMRI) acceleration, reconstructed noise in the \bx--space may show spatial heterogeneities. Pre--calibrated image--space pMRI methods, such as the Sensitivity Encoding  (SENSE) method~\cite{Pruessmann99}, allow an exact characterization of the noise model. Thus, in those methods where the reconstruction is performed in the image--space, a direct noise propagation analysis is computationally efficient, since only channel correlations need to be considered. 

Alternatively, in  pMRI methods where the reconstruction takes place into the \bk--space, e.g.: generalized autocalibrating partially parallel acquisitions (GRAPPA)~\cite{Griswold02}, the reconstruction is done coil by coil in \bk--space, using data from all the coils to reconstruct the signal from each individual coil. As a consequence, noise characterization becomes challenging, due to the introduction of correlations in the \bk--space that propagate into the \bx--space. A direct noise propagation analysis has been considered computationally infeasible due to the need to operate with very large covariance matrices~\cite{Robson08}.


An early approach to overcome this computational challenge is based on Monte-Carlo simulations \cite{Robson08}. In Monte-Carlo based methods, noise maps are obtained by repeatedly corrupting the acquired data with synthetic noise properly scaled and correlated. Unfortunately, Monte-Carlo methods are very time consuming and only provide a noisy estimation of the noise parameters. For this reason, there is significant interest in the development of analytical noise characterization in GRAPPA.

In order to avoid the need for large covariance matrices, several approximated methods have been proposed. In these methods, the problem is simplified by approximating the \bk--space image reconstruction process (which can be viewed as a convolution) as a pixel-wise multiplication in \bx--space~\cite{Breuer09,AjaMRM11}. This reformulation provides computationally efficient noise characterization by avoiding the extensive \bk--space correlations. However, the GRAPPA reconstruction is not exactly equivalent to the approximated image--space formulation, and therefore these image--space methods may generally result in errors.
Importantly, reconstructions from non-uniform \bk--space undersampling trajectories (e.g.: trajectories including a calibration region in the center of  \bk--space, the so-called ACS lines) cannot be expressed equivalently in the image--space. Although these approximated methods have shown to work properly in multiple scenarios, they will introduce errors in the presence of non-uniformly undersampled \bk--space trajectories.  Therefore, development of an exact analytical method is highly desirable.


In this work, we propose a method for exact characterization of noise in GRAPPA reconstructions. In the proposed method, the noise is propagated through the reconstruction by accounting for all the correlations in the \bk--space. The need to operate with very large covariance matrices is avoided by exploiting extensive symmetries and separability in each of the reconstruction steps. Thus, an exact and computationally efficient solution to the noise distribution in GRAPPA for every reconstructed channel is provided. 

The resulting coil-by-coil images are usually combined into a magnitude composite image using either complex \cite{Walsh00} or magnitude based \cite{Roemer90} coil combination approaches. The coil-by-coil noise estimates here provided can subsequently be combined according to the selected coil combination approach, thus characterizing noise in the final coil-combined image. In addition, this method can also be used to evaluate the noise amplification ($g$-factor) in GRAPPA reconstructions~\cite{Breuer09}.


\section{Theory}
\label{sec:theory}


\subsection{Overview of GRAPPA Acquisition and Reconstruction}

Parallel MRI methods enable accelerated acquisitions by undersampling the \bk--space, e.g.: acquiring only a fraction of the phase-encoded lines. In order to recover the missing lines, these techniques take advantage of the redundancies between data from multiple receive coils. The data acquired across all the coils ${\vec{{\bf s}} (\boldsymbol k)}=[ s_1 (\boldsymbol k) \cdots s_L (\boldsymbol k) ]$ can be modeled as \cite{AjaNoiseBook,Thunberg07}: 

\begin{equation} \label{eq:kspaceNoise}
\vec{{\bf s}}(\boldsymbol{k}) = \vec{{\bf a}}(\boldsymbol{k}) + \vec{{\bf n}}(\boldsymbol{k},{\bf \Sigma}_{k})
\end{equation}
where ${\vec{{\bf a}} (\boldsymbol k)}=[ a_1(\boldsymbol k) \cdots a_L(\boldsymbol k) ]$ is the noiseless signal in the \bk--space and ${\vec{{\bf n}}(\boldsymbol k,{\bf \Sigma}_{k})}$ is a complex  zero-mean Gaussian noise distribution with the same covariance matrix ${\bf \Sigma}_{k}$ for both the real and imaginary parts. Noise is assumed to be stationary, which implies that ${{\bf \Sigma}_{k}}$ does not depend on ${\boldsymbol k}$.

In GRAPPA, the reconstruction takes place in the \bk--space, where each point in the missing lines is computed as an interpolation from its neighborhood ${\eta (\boldsymbol k)}$ in all the coils~\cite{Griswold02}:
\begin{equation} \label{eq:GRAPPAeq_k}
s_l^R(\boldsymbol k)=\sum_{m=1}^{L} \sum_{\boldsymbol c \in \eta (\boldsymbol k)} s_m^S (\boldsymbol k- \boldsymbol c) \cdot \omega_m(l,\boldsymbol c)
\end{equation}
where $L$ is the total number of coils, $s_m^S({\boldsymbol k})$ is the sampled \bk--space signal from coil $m$, ${\omega_m(l,\boldsymbol c)}$ are the complex reconstruction weights for coil $l$, and $s_l^R({\boldsymbol k})$ is the reconstructed  \bk--space signal from coil $l$. The weights ${\omega_m(l,\boldsymbol c)}$  are usually calculated from a fully sampled low-frequency region in the \bk--space, called the Auto Calibration Signal (ACS) lines~\cite{Griswold02}.

A final composite image is obtained by merging the data from every coil into a single image, which can be done using the sum of squares (SoS) or, in a more general way, as a properly weighted linear combination \cite{Roemer90}. In this work we will use an alternative reconstruction, the linear combination proposed in \cite{Walsh00} since it allows the reconstruction to be written as a linear matrix operation.

\begin{figure}[tb]
\centering
\includegraphics[width=0.99\columnwidth]{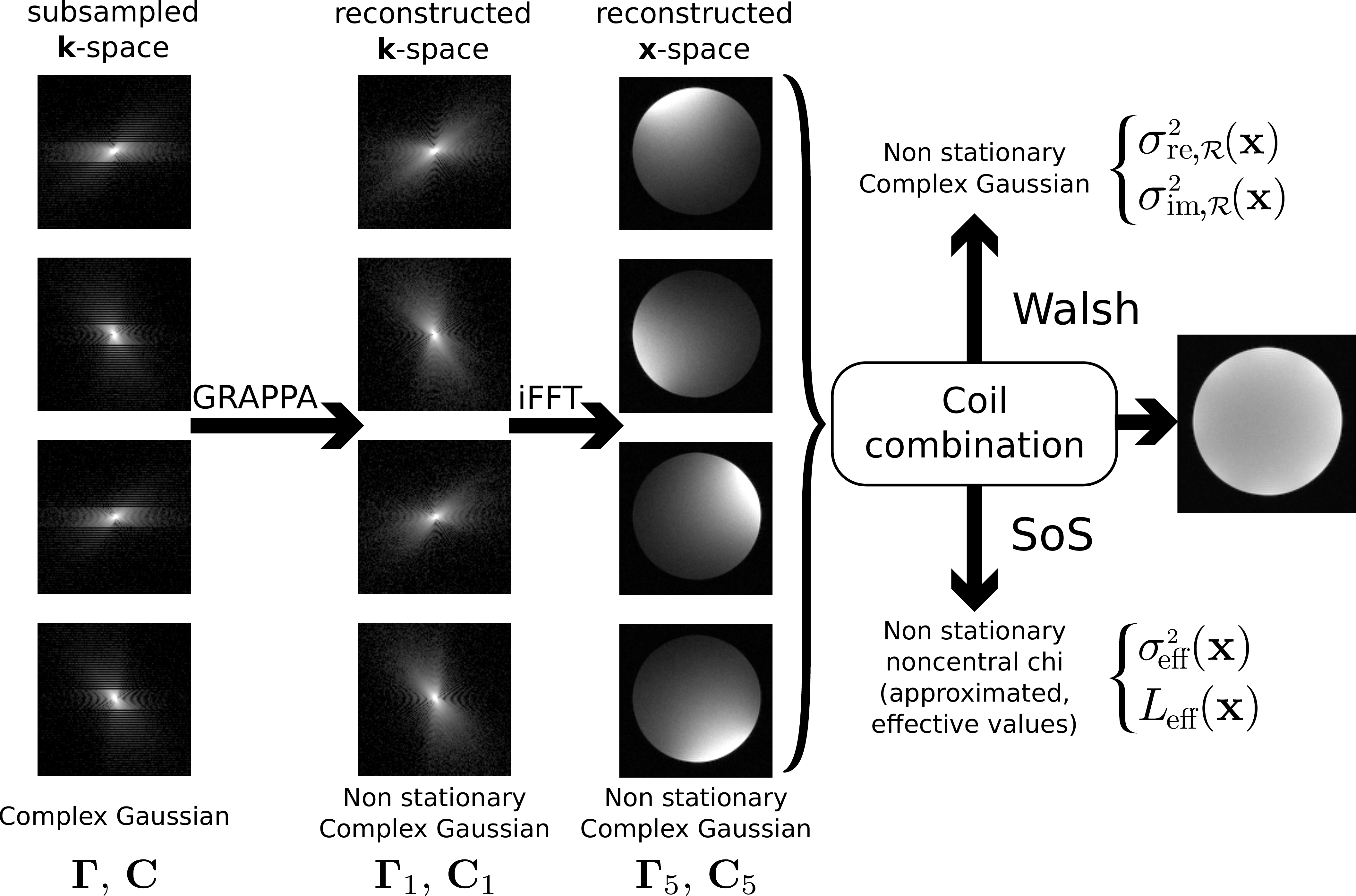}
\caption{Noise models with the main parameters related for each of the steps of the GRAPPA reconstruction pipeline. The distribution of the final composite image depends on the coil combination procedure selected.}
\label{fig:scheme}
\end{figure}

In order to characterize the noise in the final composite image, each of the steps of the GRAPPA processing must be properly characterized, see Fig.~\ref{fig:scheme}. Initially, noise in the sampled \bk-space is assumed to be a stationary complex Gaussian process. Due to the GRAPPA interpolation, noise in the reconstructed signal can also me modeled as Gaussian, but it becomes position--dependent, i.e., non-stationary. This noise is modeled by a covariance matrix containing the correlations between all the samples in the \bk--space. The size of this matrix is given by $\left[ N_p \cdot N_f \cdot L\right] \times \left[ N_p \cdot N_f \cdot L\right]$, where $N_p$ is the number of points in the phase-encoding direction and $N_f$ is the number of points in the frequency-encoding direction. Although theoretically feasible, the direct noise analysis becomes rapidly prohibitive due to the need to perform operations on very large matrices. To overcome this limitation, previously proposed methods have used models based on approximations of the reconstructed signal that simplify the problem. These approximated methods are described in the following section. 



\subsection{Image--space methods for noise characterization in GRAPPA}
\label{sec:grappa_x}

In Refs~\cite{Breuer09,AjaMRM11}, the authors propose an analysis of noise in the image--space based on rewriting eq.~\eqref{eq:GRAPPAeq_k} as a convolution 
\begin{equation} \label{eq:GRAPPAeq_k_conv}
s_l^R(\boldsymbol k)=\sum_{m=1}^{L} s_m^S (\boldsymbol k)  \circledast {\sf w}_m(l,\boldsymbol k)
\end{equation}
where ${{\sf w}_m(l,\boldsymbol k)}$ is a convolution kernel built from the reconstruction weights ${\omega_m(l,\boldsymbol c)}$ \cite{Breuer09,Brau08}. In these methods, the GRAPPA reconstruction is expressed as a pixel-wise product in the \bx--space

\begin{equation} \label{eq:GRAPPAeq_x_prod}
S_l^R(\boldsymbol x) =N_p\cdot N_f\cdot\sum_{m=1}^{L} S_m^S (\boldsymbol x)  \times W_m(l,\boldsymbol x)
\end{equation}
%
%
where $S_l^R(\boldsymbol x)$ and $S_l^S(\boldsymbol x)$ are the reconstructed and sampled signals at coil $l$-th in the \bx--space, and ${W_m(l,\boldsymbol x)}$ are the reconstruction kernels transformed into the \bx--space by the 2D-iFFT. We define the matrix ${\bf W} (\boldsymbol x)$ containing all the weights to reconstruct a pixel across all the channels:
\begin{equation} \label{eq:WeightsImage}
{\bf W} (\boldsymbol x)= \left( \begin{array}{ccc}
W_{11} (\boldsymbol x) & \cdots & W_{1L} (\boldsymbol x)\\
\vdots & \ddots & \vdots\\
W_{L1} (\boldsymbol x) & \cdots & W_{LL} (\boldsymbol x)
\end{array}\right).
\end{equation}

In \cite{AjaMRM11}, the authors consider that an undersampled GRAPPA acquisition is equivalent to zero-padding the missing lines. When the 2D-iFFT is computed, spatial stationarity is preserved, but the noise in the image--space decreases with respect to the fully-sampled case, which results in 
\begin{equation} \label{eq:sigmaXsub}
{\bf\Sigma}^{\mathcal{S}}_{x}=\frac{N_{\text{acq}}/N_p}{N_p\cdot N_f}\cdot {\bf \Sigma}_{k}=\frac{1}{R_{\text{eff}}}\cdot {\bf \Sigma}^{\text{full}}_{x}
\end{equation}
where   $N_{\text{acq}}$ is the number of acquired lines in the phase-encoding direction and $R_{\text{eff}}=\frac{N_p}{N_{\text{acq}}}$ is the effective acceleration. ${{\bf \Sigma}^{\text{full}}_{x}=\frac{1}{N_p\cdot N_f} {\bf \Sigma}_{k}}$ denotes the covariance matrix in the image--space for a fully-sampled acquisition, which preserves stationarity and spatial uncorrelation due to the orthogonality of the iFFT~\cite{AjaMRM11}.

The product in eq.~\eqref{eq:GRAPPAeq_x_prod} will introduce non-stationarity since the transformed kernels ${W_m(l,\boldsymbol x)}$ vary across the image.  The noise variance for coil $l$ in the \bx--space becomes:
\begin{equation} \label{eq:sigmaXrecX_coil}
\sigma^2_{{\mathcal{R},l}}(\boldsymbol x)=\frac{1}{R_{\text{eff}}} \left[ {\bf W}(\boldsymbol x) \cdot {\bf \Sigma}^{\text{full}}_{x} \cdot {\bf W}^H(\boldsymbol x) \right]_{ll}.
\end{equation}
In order to quantify the noise amplification due to pMRI, the $g$-factor is used:
\begin{equation} \label{eq:$g$-factor_coil}
g_l(\boldsymbol x)=\frac{\snr^{\text{full}}_{l}}{\snr^{\mathcal{R}}_{l}\cdot \sqrt{R_{\text{eff}}}}
= \frac{\sqrt{\left[ {\bf W}(\boldsymbol x) \cdot {\bf \Sigma}^{\text{full}}_{x} \cdot {\bf W}^H(\boldsymbol x) \right]_{ll}}}{\sqrt{R_{\text{eff}}} \cdot \sqrt{\left[ {\bf \Sigma}^{\text{full}}_{x} \right]_{ll}}}.
\end{equation}

Finally,  coil combination can be performed using the SoS~\cite{AjaMRM11}, producing a signal that can be approximated using a non--central $\chi$ distribution (see Fig.~\ref{fig:scheme}). Alternatively, a  complex linear coil combination can be used \cite{Walsh00}. In this case, the information from each coil is combined using a vector $\vec{{\bf m}} (\boldsymbol x)=\left[m_1 (\boldsymbol x) \cdots m_L(\boldsymbol x) \right]^T$:
\begin{equation} \label{eq:SMF}
S_T(\boldsymbol x)=\vec{{\bf m}}(\boldsymbol x) \cdot \vec{{\bf S}^R} (\boldsymbol x) = \sum_{l=1}^L m_l(\boldsymbol x) S_l^R(\boldsymbol x).
\end{equation}
For complex linear coil combination, the noise in the composite image is  complex Gaussian , with variance and $g$-factor as follows
%
%
\begin{equation}
\sigma_{T}^2(\boldsymbol x)=\frac{1}{R_{\text{eff}}}\ \vec{{\bf m}}(\boldsymbol x) \cdot {\bf W}(\boldsymbol x) \cdot {\bf \Sigma}^{\text{full}}_{x} \cdot {\bf W}^H(\boldsymbol x) \cdot \vec{{\bf m}}^H(\boldsymbol x),\label{eq:sigmaXrecX_combined}
\end{equation}
\begin{equation}
g_T(\boldsymbol x)=\frac{\sqrt{\vec{{\bf m}}(\boldsymbol x) \cdot {\bf W}(\boldsymbol x) \cdot {\bf \Sigma}^{\text{full}}_{x} \cdot {\bf W}^H(\boldsymbol x) \cdot \vec{{\bf m}}^H(\boldsymbol x)}}{\sqrt{R_{\text{eff}}} \cdot \sqrt{\vec{{\bf m}}(\boldsymbol x) \cdot {\bf \Sigma}^{\text{full}}_{x} \cdot \vec{{\bf m}}^H(\boldsymbol x)}}.
\label{eq:$g$-factor_combined}
\end{equation}

This equivalence between the reconstruction in both spaces holds true when the undersampling pattern is uniform across the \bk--space, as in the case when the weights are obtained from a separate pre-scan. However, if the ACS region is acquired within the image, or a Variable Density (VD) GRAPPA acquisition scheme \cite{Heidemann07} is used, it is not possible to reconstruct the whole \bk--space with a single kernel using Eq.~\eqref{eq:GRAPPAeq_k_conv}. Ref~\cite{Breuer09} proposes a model based on splitting the \bk--space into multiple regions, where each region is uniformly undersampled. Each  \bk--space region is then reconstructed with its own kernel, resulting in the following expressions for the variance and $g$-factor in the final image:
\begin{equation} \label{eq:sigmaXrecX_multiKernel_combined}
\sigma_{T}^2(\boldsymbol x)=\sum_{m=1}^{M} \frac{f_m}{R_m} \vec{{\bf m}}(\boldsymbol x) \cdot {\bf W}_m(\boldsymbol x) \cdot {\bf \Sigma}^{\text{full}}_{x} \cdot {\bf W}_m^H(\boldsymbol x) \cdot \vec{{\bf m}}^H(\boldsymbol x),
\end{equation}
\begin{equation} \label{eq:$g$-factor_multiKernel_combined}
g_T(\boldsymbol x)=\sum_{m=1}^{M} \frac{f_m}{R_m} \frac{\sqrt{\vec{{\bf m}}(\boldsymbol x) \cdot {\bf W}_m(\boldsymbol x) \cdot {\bf \Sigma}^{\text{full}}_{x} \cdot {\bf W}_m^H(\boldsymbol x) \cdot \vec{{\bf m}}^H(\boldsymbol x)}}{\sqrt{R_{\text{eff}}} \cdot \sqrt{\vec{{\bf m}} (\boldsymbol x) \cdot {\bf \Sigma}^{\text{full}}_{x} \cdot \vec{{\bf m}}^H (\boldsymbol x)}} 
\end{equation}
where $M$ is the total number of regions, ${\bf W}_m(\boldsymbol x)$ is the kernel used in the region $m$, $f_m$ is the fraction of \bk--space in a for that region. 

\subsubsection{\underline{Limitations of image--space methods}} 
Although image--space methods have shown to be useful in practical situations, they have fundamental limitations, particularly in the presence of a calibration region or non-uniform undersampling. One of the advantages of GRAPPA is precisely its ability to accomodate variable density undersampling schemes, where regions with different acceleration factors need to be reconstructed using different kernels. In this scenario assuming that the different \bk--space regions are statistically uncorrelated, image--space methods reconstruct the coil images in several steps. First, the \bk--space is split into the differently undersampled regions. Then, each region is transformed into the image--space and subsequently reconstructed as a product with the 2D-iFFT of its kernel, generating an image per region. Finally, these region-specific images are added to obtain the final coil image.

This methodology produces two sources of error: first, the reconstruction in the \bx--space is not exactly equivalent to the reconstruction in the \bk--space. When a kernel is applied to a filtered region, the support of the output overlaps with the adjacent regions, introducing a residual error; second, when different acceleration factors are used, the acquired lines in the boundaries that separate every pair of regions are used to reconstruct both regions. Consequently, the two regions will be correlated. For these reasons, the image--space analysis (including equations \ref{eq:sigmaXrecX_multiKernel_combined} and \ref{eq:$g$-factor_multiKernel_combined}) is not fully correct.


\subsection{\bk--space method for noise characterization in GRAPPA}
\label{sec:grappa_x}

To overcome the limitations of image--space methods, an efficient \bk--space noise propagation analysis is proposed. It avoids operating with very large matrices by exploiting symmetries and separability in the covariance matrices. This analysis is exact under the assumptions of stationarity and uncorrelation in the original undersampled \bk--space acquisition. Furthermore, the analysis is exact also for more complex  \bk--space undersampling, including variable density methods~\cite{Heidemann07}.

For the sake of simplicity, our analysis considers: (1) GRAPPA weights are non-stochastic, i.e., they are independent of the noise realization. This is strictly true only if they are estimated from an independent acquisition; however we expect this assumption to be a good approximation even in self-calibrated acquisitions, given the typical overdetermination of GRAPPA weight estimation; (2) acquired points in the \bk--space are IID.

For every acquired point in the \bk--space we obtain a complex vector $\vec{\bf p}=\vec{\bf x}+i\cdot \vec{\bf y}$ containing the measures for all the coils\footnote{$\vec{p}$ is the stack of componentes $s_m^S (\boldsymbol k), 1\leq m\leq L$ in Equation \ref{eq:GRAPPAeq_k_conv}. We hereafter use this symbol for simplicity.}, having a $L$-variate complex normal distribution for every point characterized by three parameters~\cite{Goodman63}:

\begin{equation} \label{eq:complexMVparameters}
\begin{split}
\vec{\boldsymbol \mu_p}&=E\{\vec{\bf p}\}, \\
{\bf \Gamma}&=E\{(\vec{\bf p}-\vec{\boldsymbol \mu_p})(\vec{\bf p}-\vec{\boldsymbol \mu_p})^H\} \\&= {\bf \Sigma}_{\text{re-re}} + {\bf \Sigma}_{\text{im-im}} +i \cdot \left( {\bf \Sigma}_{\text{im-re}} -{\bf \Sigma}_{\text{re-im}} \right), \\
{\bf C}&=E\{(\vec{\bf p}-\vec{\boldsymbol \mu_p})(\vec{\bf p}-\vec{\boldsymbol \mu_p})^T\} \\&= {\bf \Sigma}_{\text{re-re}} - {\bf \Sigma}_{\text{im-im}} +i \cdot \left( {\bf \Sigma}_{\text{im-re}} + {\bf \Sigma}_{\text{re-im}} \right) 
\end{split}
\end{equation}
where ${\bf \Sigma}_{ \left\lbrace \text{re,im} \right\rbrace - \left\lbrace \text{re,im}\right\rbrace}$ are the covariance matrices between real and imaginary components (assuming null mean). Note that the  image--space analysis methods described above were restricted to ${{\bf \Sigma}_{\text{im-re}}={\bf \Sigma}_{\text{re-im}}=0}$ and ${\bf \Sigma}_{\text{re-re}}={\bf \Sigma}_{\text{im-im}}={\bf \Sigma}$ and ${{\bf \Gamma}=2\cdot {\bf \Sigma}}$. However, if there is a correlation between real and imaginary components across different channels \cite{Roemer90}, ${{\bf C}\neq 0}$ and correlations must be taken into account.

\subsubsection{\underline{\bk--space Interpolation}}
\label{sec:grappa_interpolation} 
expressed for every point as
\begin{equation} \label{eq:GrappaRecPoint}
\vec{\bf p}(\boldsymbol k)=\sum_{\vec{{\bf q}_k} \in \eta (\boldsymbol k)} \boldsymbol{\W}^{\vec{\bf p}(\boldsymbol k)}_{\vec{{\bf q}_k}}  \cdot \vec{{\bf q}_k} ,
\end{equation}
where ${\vec{\bf p}(\boldsymbol k)}$ contains the \bk--space value for the point ${\boldsymbol k}$ and the vector ${\vec{\bf q}_k=\left[ q_1 \cdots q_L \right]}$ contains the \bk--space values for all the channels of the points contained in its neighborhood ${\eta(\boldsymbol k)}$. $\boldsymbol{\W}^{\vec{\bf p}(\boldsymbol k)}_{\vec{{\bf q}_k}}$ is an ${L\times L}$ matrix in which the $l$-th row contains the GRAPPA weights 
${\vec{\boldsymbol{\omega}}(l,\vec{{\bf q}_k}) = [\omega_1(l,\vec{{\bf q}_k})	\ldots \omega_L(l,\vec{{\bf q}_k})]}$ 
from Eq.~\ref{eq:GRAPPAeq_k}	 associated to the kernel point ${\vec{{\bf }q_k}}$ to reconstruct the \bk--space value in the $l$-th coil. 

The GRAPPA interpolation introduces correlations in the reconstructed \bk--space. The correlation between two arbitrary points in the reconstructed \bk-space is
\begin{equation} \label{eq:GammaGrappa}
\begin{split}
{\bf \Gamma}_1 & = E\{\vec{\bf p}_i \cdot \vec{\bf p}_j^H\} \\
& = E\left\{\left(\sum_{\vec{{\bf q}_{k_i}} \in \eta_i} \boldsymbol{\W}_{\vec{{\bf q}_{k_i}}}^{\vec{{\bf p}_i}} \cdot \vec{{\bf q}_{k_i}} \right) \left(\sum_{\vec{{\bf q}_{k_j}} \in \eta_j} \boldsymbol{\W}_{\vec{{\bf q}_{k_j}}}^{\vec{{\bf p}_j}} \cdot \vec{{\bf q}_{k_j}} \right)^H\right\} \\
& = \sum_{\vec{{\bf q}_{k_i}} \in \eta_i} \sum_{\vec{{\bf q}_{k_j}} \in \eta_j} \boldsymbol{\W}_{\vec{{\bf q}_{k_i}}}^{\vec{{\bf p}_i}} \cdot E\left\{ \vec{{\bf q}_{k_i}} \cdot \vec{{\bf q}_{k_j}}^H\right\} \cdot {\boldsymbol{\W}_{\vec{{\bf q}_{k_j}}}^{\vec{{\bf p}_j}}}^H  \\
& = \sum_{\vec{{\bf q}_{k}} \in \left( \eta_i \cap \eta_j \right)}  \boldsymbol{\W}_{\vec{{\bf q}_{k}}}^{\vec{{\bf p}_i}} \cdot {\bf \Gamma} \cdot {\boldsymbol{\W}_{\vec{{\bf q}_{k}}}^{\vec{{\bf p}_j}}}^H  \\
\end{split}
\end{equation}
and equivalently
\begin{equation} \label{eq:CGrappa}
{\bf C}_1=E\{\vec{{\bf p}_i} \cdot \vec{{\bf p}_j}^T\}=\sum_{\vec{{\bf q}_{k}} \in \left( \eta_i \cap \eta_j \right)}  \boldsymbol{\W}_{\vec{{\bf q}_{k}}}^{\vec{{\bf p}_i}} \cdot {{\bf \Gamma}} \cdot {\boldsymbol{\W}_{\vec{{\bf q}_{k}}}^{\vec{{\bf p}_j}}}^T.
\end{equation}

\begin{figure}[tb]
\centering
\includegraphics[width=\columnwidth]{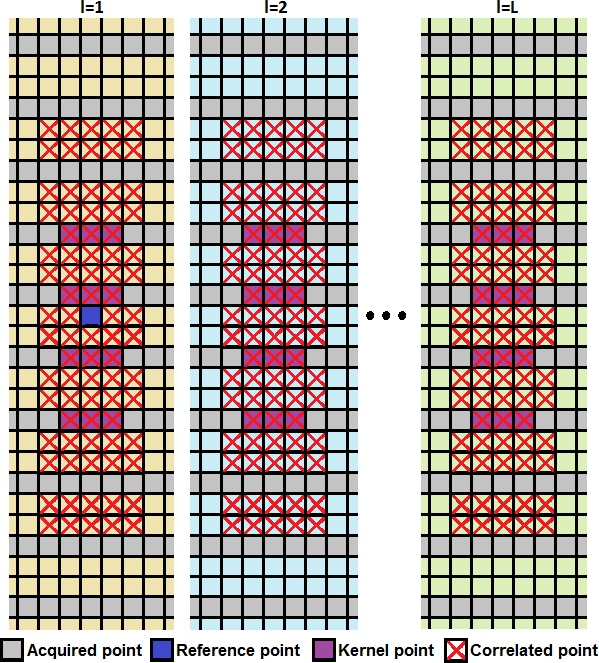}
\caption{Example of correlations between a point and its neighborhood for the case of 2D-GRAPPA with acceleration factor 2 and kernel size $4\times 3$. A missing point in \bk--space is estimated as an interpolation of its neighborhood. Consequently, it will only be correlated with a reduced number of points in \bk--space consisting of the points in its neighborhod and those points whose neighborhood overlaps with its own neighborhood.}
\label{fig:correlations}
\end{figure}

This implies that a point in the reconstructed \bk--space correlates with all the points with overlapping neighborhoods. The points in the acquired lines are left untouched, and thus their kernel only contains one point (themselves), so $\boldsymbol{\W}_{\vec{{\bf q}_{k}}}^{\vec{{\bf p}}}$ is the identity matrix for $\vec{\bf p}=\vec{{\bf q}_{k}}$ and the null matrix otherwise. Fig.~\ref{fig:correlations} shows an example of the correlations introduced by GRAPPA reconstruction.

If we define the GRAPPA kernel size as ${ \left[K_p,K_f\right]}$, a reconstructed point correlates in the frequency-encoding direction with points in ${2K_f-1}$ columns, where a column refers to the data for a particular frequency-encoding value and a row refers to the data for a particular phase-encoding value. In the phase-encoding direction, the number of points with which a reconstructed point correlates  depends on the row since the kernel may vary from row to row. Let us now stack a set of consecutive ${2K_f-1}$ columns into a vector; the column in position $K_f$ in the stack will be referred to as reference column. If we now stack $L$ such vectors (one from each coil), we construct a $N_p\cdot\left(2 K_f-1 \right )\cdot L$-component vector, with correlation matrix defined as follows (for simplicity we only keep track of the evolution of ${\bf \Gamma}$):
%
%
%
\begin{equation} \label{eq:GammaInter}
{\bf \Gamma}_2 = \left( \begin{array}{cccc}
{\bf B}_{11} & {\bf B}_{12} & \cdots & {\bf B}_{1L}\\
{\bf B}_{21} & {\bf B}_{22} & \cdots & {\bf B}_{2L}\\
\vdots & \vdots & \ddots & \vdots\\
{\bf B}_{L1} & {\bf B}_{L2} & \cdots & {\bf B}_{LL}
\end{array}\right)
\end{equation}
where ${\bf B}_{ij}$ contains the correlations of the reference column and its $2K_f-2$ neighboring columns in the $i$-th coil and this equivalent vector in the $j$-th coil. This matrix presents Hermitian symmetry, ${\bf B}_{ij}={\bf B}_{ji}^H$; thus, we only need to compute ${L\cdot (L+1)/2}$ blocks.

Since GRAPPA uses the same kernel for all the reconstructed points in a row, under the stationarity assumption, the correlation between a reference column and its surrounding columns is the same, independently of the column picked for reference. This gives rise to a block-Toeplitz structure for each of the blocks ${\bf B}_{ij}$ in eq.~\eqref{eq:GammaInter}. Specifically, if the reference column has index $K_f$, its left hand side columns are indexed within the interval $1\leq i \leq K_f-1$ and its right hand side columns are indexed within  $ K_f+1 \leq i \leq  2K_f-1$, then the first column of ${\bf B}_{ij}$ is  defined as
\begin{equation} \label{eq:FirstColumnOfB}
\left [\begin{array}{ccccccc}
{\bf b}^{0}_{ij} & {\bf b}^{1}_{ij} & \cdots & {\bf b}^{K_f-1}_{ij} & \bf \overline{0} & \cdots & \bf \overline{0} 
\end{array} \right ]^T
\end{equation}
where ${\bf b}^{m}_{ij}$ is the $N_p\times N_p$ covariance matrix of two columns, the subtraction of the indices of which equals $m$, and $\bf \overline{0}$ denotes a null matrix of dimensions $N_p\times N_p$. These covariances are obtained by selecting the appropriate components of ${\bf \Gamma}_1$ in eq.~\eqref{eq:GammaGrappa}. As for the first row, 
\begin{equation} \label{eq:FirstRowOfB}
\left [\begin{array}{ccccccc}
{\bf b}^0_{ij} & {\bf b}^{-1}_{ij} & \cdots & {\bf b}^{1-K_f}_{ij} & \bf \overline{0} & \cdots & \bf \overline{0} 
\end{array} \right ].
\end{equation}
%
%
%
%
Notice that we only need to compute ${2K_f-1}$ sub-blocks due to the block-Toeplitz structure as well as the presence of null correlations known beforehand. Unfortunately no conjugate symmetry generally holds for these sub-blocks, but only for $i=j$. Thefore, the overall number of sub-blocks needed to build \ref{eq:GammaInter} is at most ${(2K_f-1) \cdot L\cdot (L+1)/2}$.


\subsubsection{\underline{Column iFFT}}
\label{sec:grappa_col_iFFT} 

After reconstructing the \bk--space, data are transformed into the image--space by a 2D-iFFT, which can be decomposed into two 1D-iFFT. Computing the 1D-iFFT along the phase-encoding direction preserves the number of points with which a reference point correlates  across the frequency-encoding direction.

The 1D-iFFT can be expressed as a matrix operation with matrix ${{\bf F}_I}$. For our vector stack, the 1D-iFFT of each column for each coil can be obtained by the product with: 
\begin{equation} \label{eq:ComposeiFFT}
\boldsymbol{\F}_I = \left( \begin{array}{cccc}
{\bf F}_I & \bf \overline{0} & \cdots & \bf \overline{0}\\
\bf \overline{0} & {\bf F}_I & \cdots & \bf \overline{0}\\
\vdots & \vdots & \ddots & \vdots\\
\bf \overline{0} & \bf \overline{0} & \cdots & {\bf F}_I
\end{array}\right).
\end{equation}
The correlation matrix of the transformed vector is straightforwardly obtained as  
%
\begin{equation} \label{eq:GammaHybridCol}
{\bf \Gamma}_3 = \boldsymbol{\F}_I \cdot {\bf \Gamma}_2 \cdot \boldsymbol{\F}_I^H = \left( \begin{array}{cccc}
{\bf D}_{11} & {\bf D}_{12} & \cdots & {\bf D}_{1L}\\
{\bf D}_{21} & {\bf D}_{22} & \cdots & {\bf D}_{2L}\\
\vdots & \vdots & \ddots & \vdots\\
{\bf D}_{L1} & {\bf D}_{L2} & \cdots & {\bf D}_{LL}
\end{array}\right)
\end{equation}
where each ${\bf D}_{ij}$ shows the same block-Toeplitz structure as ${\bf B}_{ij}$, i.e., just a replacement of $\bf b$ by $\bf d$ is needed in expressions  \ref{eq:FirstColumnOfB}  and \ref{eq:FirstRowOfB} to build ${\bf D}_{ij}$, 
%
%
due to the fact that
\begin{equation} \label{eq:FFTblockCol}
{\bf d}^m_{ij} = {\bf F}_I \cdot {\bf b}^m_{ij} \cdot {\bf F}_I^H.
\end{equation}


Once again, the number of sub-blocks  ${\bf d}_{ij}$ needed to build  ${\bf D}_{ij}$ in \ref{eq:GammaHybridCol} is ${(2K_f-1) \cdot L\cdot (L+1)/2}$, i.e., it coincides with the number needed to build \ref{eq:GammaInter} because of equality \ref{eq:FFTblockCol}.
 

\subsubsection{\underline{Row iFFT}}
\label{sec:grappa_row_iFFT}

To complete the 2D-iFFT a second 1D-iFFT is done along the row dimension. 
The correlation within each row suffices to obtain the noise maps. It is worth mentioning that this analysis generalizes to calculate the correlations between any pair of rows directly.  

First, we create a column vector by stacking a selected row (set up as a column) in \bk-space with the same rows in the other coils; this vector will have $N_f \cdot L$ components. The correlation matrix of this vector can be defined as 
\begin{equation} \label{eq:GammaHybridRow}
{\bf \Gamma}_4 = \left( \begin{array}{cccc}
{\bf G}_{11} & {\bf G}_{12} & \cdots & {\bf G}_{1L}\\
{\bf G}_{21} & {\bf G}_{22} & \cdots & {\bf G}_{2L}\\
\vdots & \vdots & \ddots & \vdots\\
{\bf G}_{L1} & {\bf G}_{L2} & \cdots & {\bf G}_{LL}
\end{array}\right)
\end{equation}
where every block ${\bf G}_{lm}$ has dimensions ${N_f \times N_f}$. and contains the (cross-)correlations of the selected row in coils $l$ and $m$ within the hybrid space ${(k_f,y)}$. These correlations are obtained by selecting the appropriate components in \ref{eq:GammaHybridCol}. Assuming GRAPPA performs a circular interpolation, this will result in a cyclical structure along the frequency-encoding direction. Specifically, the first column of block ${\bf G}_{ij}$ is defined as
\begin{equation} \label{eq:FistRowGij}
%
%
\vec{{\bf g}_{ij}}=\left [ g^0_{ij}, \  g^{1}_{ij}, \  \cdots, \  g^{K_f-1}_{ij}, \  0, \ \cdots, \  0, \   g^{1-K_f}_{ij}, \  \cdots, \   g^{-1}_{ij}\right]
\end{equation}
with $N_f-(2K_f-1)$ zeroes and values $g^{m}_{ij}$ are taken from the components in sub-block ${\bf d}^{m}_{ij}$ in \ref{eq:FFTblockCol} that correspond to the selected row. Then, the $j$-th row of ${\bf G}_{ij}$, $2\leq j \leq N_f-1$, is obtained as a rightward circular shift of the row $j-1$, which results in the blocks ${\bf G}_{lm}$ being circulant matrices.

Computing the row 1D-iFFT provides a correlation matrix given by:
\begin{equation} \label{eq:GammaCoilsImage}
{\bf \Gamma}_5 = \boldsymbol{\F}_I \cdot {\bf \Gamma}_4 \cdot \boldsymbol{\F}_I^H = \left( \begin{array}{cccc}
{\bf H}_{11} & {\bf H}_{12} & \cdots & {\bf H}_{1L}\\
{\bf H}_{21} & {\bf H}_{22} & \cdots & {\bf H}_{2L}\\
\vdots & \vdots & \ddots & \vdots\\
{\bf H}_{L1} & {\bf H}_{L2} & \cdots & {\bf H}_{LL}
\end{array}\right)
\end{equation}
where the blocks ${\bf H}_{ij}$ are:
\begin{equation} \label{eq:FFTblockRow}
{\bf H}_{ij} = {\bf F_I} \cdot {\bf G}_{ij} \cdot {\bf F_I}^H.
\end{equation}

Taking into account that circulant matrices are diagonalized by the FT \cite{DavisBook}, eq.~\eqref{eq:FFTblockRow} is simplified to:
\begin{equation} \label{eq:FFTblockRowCirculant}
{\bf H}_{ij} = {\bf F}_I \cdot {\bf G}_{ij} \cdot {\bf F}_I^H = \text{diag}({\bf F}_I \cdot \vec{{\bf g}_{ij} }).
\end{equation}
Again, we have Hermitian symmetry, so we only need to compute ${\left[ N_f \cdot L \cdot (L+1)/2 \right]}$ 1D-iFFT for this step.

\subsubsection{\underline{Coil combination}}
\label{sec:coil_combination}

In order to characterize the distribution in every final reconstructed pixel, we first need to obtain the correlation matrices for each pixel $\vec{\bf p}$ along the coil dimension. The matrix ${\bf \Gamma}_5$ contains the correlations between each point in a row and all the points in that row in all the coils. The 1D-iFFT in the frequency-encoding direction introduces non-stationarity along this dimension, so we need to proceed pixel by pixel and extract the correlation matrices ${\bf \Gamma}_6$ and ${\bf C}_6$ for a selected position, which consists of picking the corresponding entry in the diagonal of ${\bf \Gamma}_5$ or ${\bf C}_5$.

If the complex images are linearly combined as in Eq.~\eqref{eq:SMF}, the final image preserves the Gaussian behavior (see Fig.~\ref{fig:scheme}), although it presents spatial correlations and non-stationarity. For every pixel in the composite image, we can define 
\begin{equation} \label{eq:GammaSMF}
{\Gamma}_7(\boldsymbol x)=\vec{{\bf m}}_l(\boldsymbol x) \cdot {\bf \Gamma}_6(\boldsymbol x) \cdot \vec{{\bf m}}_l^H(\boldsymbol x).
\end{equation}
The variance of noise for the real and imaginary components can be calculated by:
\begin{equation} \label{eqn:finalSigma}
\begin{split}
\sigma^2_{\text{re},{\mathcal{R}}}(\boldsymbol x)=\frac{1}{2} \text{Re}\left\lbrace {\Gamma}_7 + {C}_7 \right\rbrace,  \\
\sigma^2_{\text{im},{\mathcal{R}}}(\boldsymbol x)=\frac{1}{2} \text{Re}\left\lbrace {\Gamma}_7 - {C}_7 \right\rbrace.	
\end{split}
\end{equation}
Note that correlation between real and imaginary components for a pixel can exist, which are computed as:
\begin{equation} \label{eqn:finalSigmaRealImag}
\begin{split}
\sigma_{\text{re-im},\mathcal{R}}^2(\boldsymbol x)=\frac{1}{2} \text{Im}\left\lbrace C_7 - \Gamma_7 \right\rbrace,  \\
\sigma_{\text{im-re},\mathcal{R}}^2(\boldsymbol x)=\frac{1}{2} \text{Im}\left\lbrace \Gamma_7 + C_7 \right\rbrace .	
\end{split}
\end{equation}
Finally, the $g$-factor map is derived from the previous equation defining a {\em mean} variance ${\sigma_{\mathcal{R}}^2(\boldsymbol x)}$
\begin{equation} \label{eqn:meanVar}
\sigma_{\mathcal{R}}^2(\boldsymbol x)=\frac{ \sigma_{\text{re},\mathcal{R}}^2(\boldsymbol x) + \sigma_{\text{im},\mathcal{R}}^2(\boldsymbol x)}{2},
\end{equation}
\begin{equation} \label{eq:$g$-factor_final}
g_{\mathcal{R}}(\boldsymbol x)=\frac{\sqrt{\sigma_{\text{re},\mathcal{R}}^2(\boldsymbol x) + \sigma_{\text{im},\mathcal{R}}^2(\boldsymbol x)}}{\sqrt{R_{\text{eff}}} \cdot \sqrt{\vec{{\bf m}}(\boldsymbol x) \cdot \left( {\bf \Sigma}^{\mathcal{R}}_{x,\text{re-re}} + {\bf \Sigma}^{\text{full}}_{x,\text{im-im}} \right) \cdot \vec{{\bf m}}^H(\boldsymbol x)}} .
\end{equation}

Although only results for the linear combination in Eq.~\eqref{eq:SMF} have been studied here, the application to other coil combination methods (e.g.: SoS) is straightforward once the covariance matrices ${\bf \Gamma}_6$ and ${\bf C}_6$ are calculated.

\subsubsection{\underline{Summary of the procedure}}

The procedure to obtain ${\bf \Gamma}$ matrices can be summarized as follows ($\bf C$ matrices are obtained similarly): 
\begin{enumerate}

\item Calculate ${\bf b}^m_{ij}$; the number of such sub-blocks is ${(2K_f-1) \cdot L\cdot (L+1)/2}$, and their dimension is $N_p\times N_p$. Each entry in these sub-blocks is obtained from \ref{eq:GammaGrappa}, by choosing the appropriate components. 

\item Calculate ${\bf d}^m_{ij}$ using \ref{eq:FFTblockCol}.

\item Calculate $\vec{{\bf g}_{ij}}$ in \ref{eq:FistRowGij} by choosing the appropriate components of ${\bf d}^m_{ij}$. 

\item Calculate ${\bf H}_{ij}$ using \ref{eq:FFTblockRowCirculant}.

\item Create matrix ${\bf \Gamma}_6$ by selecting a pixel index in the row, say $l$, then $[{\bf \Gamma}_6]_{(i,j)}={\bf H}_{i,j}(l,l)$. 

\item Apply equations \ref{eq:GammaSMF}, \ref{eqn:meanVar} and \ref{eq:$g$-factor_final} to obtain pixel-wise noise characterization.

\end{enumerate}

\begin{figure*}[tb]
\centering
\includegraphics[width=1\textwidth]{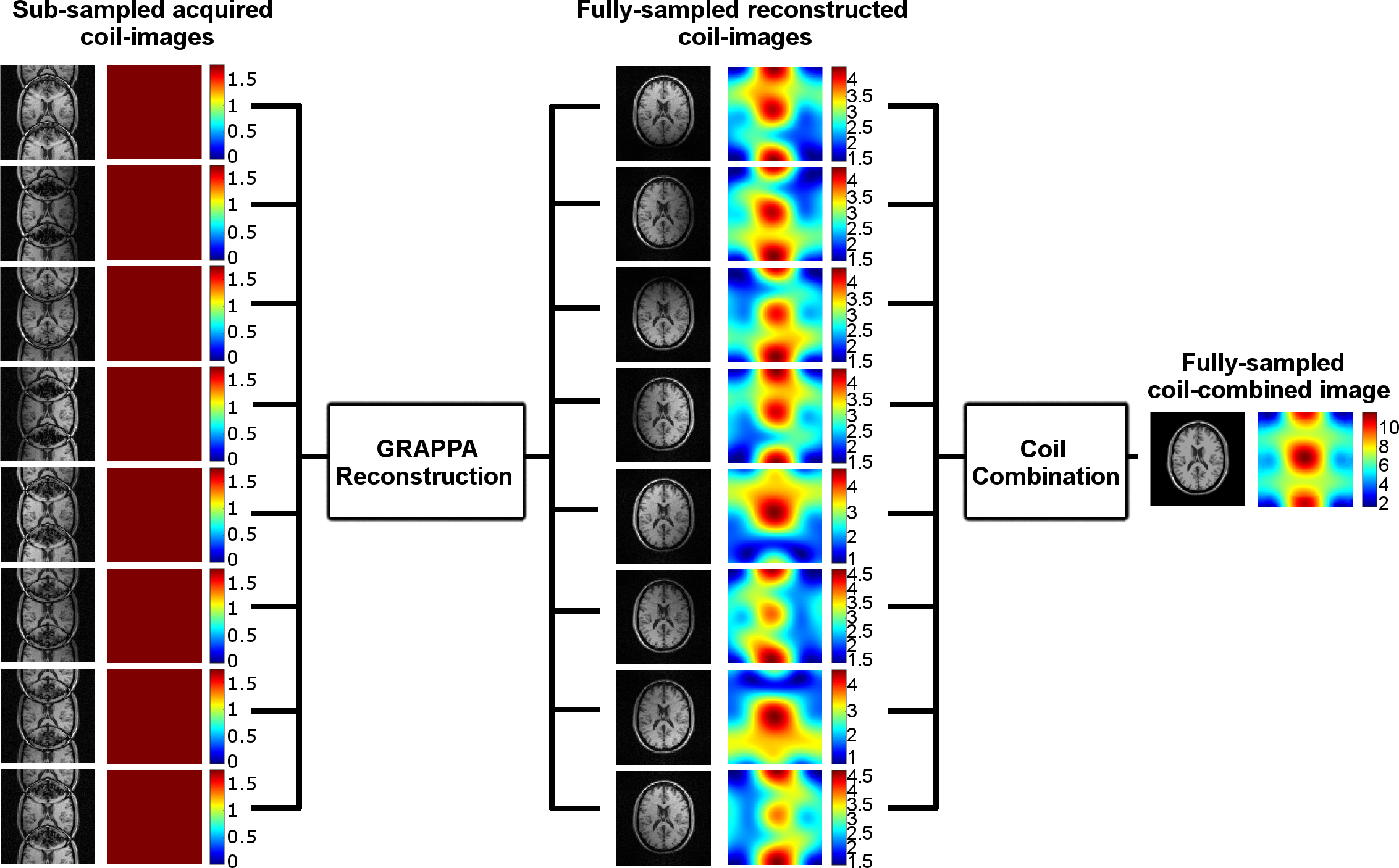}
\caption{Pipeline of GRAPPA reconstruction. Undersampled images are acquired in multiple channels, showing stationary noise. GRAPPA reconstructs a complex image for every channel, introducing noise heterogeneity. Finally, complex coil images are combined linearly and the final image also shows a non-stationary noise.}
\label{fig:propagation}
\end{figure*}

Fig.~\ref{fig:propagation} shows an outline of the noise propagation throughout the entire GRAPPA reconstruction. The image is acquired through different channels and noise can be considered stationary across the image for each channel, although correlations between channels can be present. GRAPPA then reconstructs an image for each channel interpolating the missing \bk--space areas, an operation that introduces spatial non-stationarity in the image--space. Last, the different channels need to be combined in order to provide the final image, which will also show spatial non-stationarity.

\subsection{Computational complexity}

As mentioned above, a naive \bk--space approach would require large amounts of memory to store ${\left[ N_p\cdot N_f \cdot L\right]\times \left[  N_p\cdot N_f \cdot L \right]}$ complex matrices containing the correlations between all the points across all the channels. Following this approach, ${O\left( \left( N_p\cdot N_f \cdot L\right)^3 \right)}$ complex multiplications would be performed to compute the noise maps.

By taking advantage of the separability into independent blocks, our method  achieves the storage of all the correlations in ${L \cdot (L-1) \cdot (2K_f-1)}$ complex matrices (${\bf \Gamma}$ and ${\bf C}$) of size ${ N_p\times N_p}$. Since these blocks can be processed independently, it is possible to operate with them sequentially and to avoid the need to store all of their data simultaneously. Regarding the number of operations, it is necessary to compute ${(2K_f-1)\cdot L \cdot (L+1)/2}$ 2D-iFFT for the phase encoding direction and ${N_f\cdot L \cdot (L+1)/2}$ 1D-iFFT for the frequency encoding direction. The total number of required complex multiplications is therefore ${(4K_f-1) \cdot L \cdot (L+1)/2 \cdot  O\left( N_f^2 \cdot \log(N_f) \right)}$.

For comparison with our proposed method, the direct analysis in the image--space \cite{Breuer09,AjaMRM11} can highly reduce the memory requirements. Since the channel correlations in the acquired sub-sampled images are spatially stationary across the image, it suffices to store the ${L\times L}$ complex correlation matrices for a single pixel. Thus, the main memory requirement arises from the need to store the ${N_f \times N_p}$ reconstruction kernels ${W_m(l,\boldsymbol x)}$ transformed into the image domain. Further, $M \cdot L^2 \cdot O\left( N_f^2\cdot \log(N_f) \right)$ complex multiplications are performed to compute the 2D-iFFT of the kernels for the $M$ differently undersampled regions, and ${M\cdot L^2\cdot  N_p\cdot N_f}$ complex multiplications are performed to compute the noise maps for all the channels.


\section{Methods} 

\begin{figure}[tb]
\centering
\includegraphics[width=1\columnwidth]{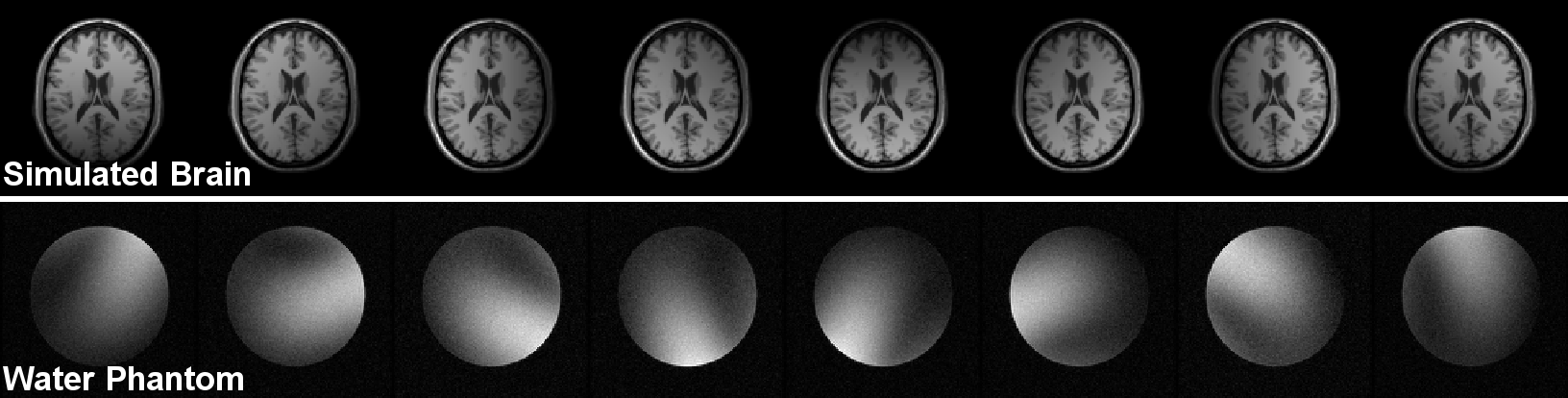}
\caption{Data sets for the experiments.}
\label{fig:datasets}
\end{figure}

Two data sets are considered for the experiments (see Fig.~\ref{fig:datasets}):

\begin{itemize}
	\item {\em Simulated brain dataset:} a reference axial brain MR image was obtained from the BrainWeb database~\cite{Collins98}. This is a T1-weighted image, with intensity non-uniformity set to INU=0\%, slice thickness=1 mm and intensity range normalized to [0,255]. An $L$-coil acquisition was simulated by modulating the image using artificial sensitivity maps coded for each coil \cite{AjaNoiseBook,AjaSENSE}. The noise-free coil images were transformed into the  \bk--space and corrupted with synthetic Gaussian noise characterized by the matrices $\Gamma_k$ and $C_k$ with SNR=30 for each coil, and the correlation coefficient between coils was set to $\rho=0.1$.  For statistical purposes, 4000 realizations of each image were used. 
	\item {\em Water phantom acquisition:} 100 realizations of the same fully-encoded slice of a water phantom doped with $3.3685$ g/L of nickel chloride hydrate (NiCl$_2$-6H$_2$O) and $2.4$g/L of sodium chloride (NaCl), scanned in an 8-channel head coil on a 3.0T scanner (MR750, GE Healthcare, Waukesha, WI). Acquisition parameters included: TE/TR=2.0/11.8ms, flip angle=3$^\circ$, FOV=220x220mm$^2$, matrix size=128x128, slice thickness=3mm, bandwidth=$\pm$62.5KHz, total scan time=641.2 seconds. 
In order to ensure steady-state, we acquired 200 realizations and discarded the first 100. Also, we corrected for B$_0$ field drift related phase variations by pre-processing step that estimated the phase-shift between realizations from the center of the \bk--space as a cubic function of time and removed it afterwards. 

\end{itemize}

From these data sets, three experiments were performed:


\begin{enumerate}
	\item First, we studied the case of uniform undersampling patterns. The \bk--space data corresponding to the first realization in each data set was subsampled with acceleration factor $R=2$ and $32$ ACS lines. The GRAPPA kernel and the combination vector \cite{Walsh00} are estimated from the ACS lines from the first realization. For each realization, the \bk--space data were undersampled uniformly without including ACS lines. Next, each realization was reconstructed using the same kernel (from the first realization) and linearly combined with the previously computed coil combination vector. Note that, in this case, the assumptions underlying image--space based methods are correct, and therefore image--space methods should be equivalent to the proposed  \bk--space method for this first case. 
	\item Second, we considered the case of non-uniform subsampling patterns due to the presence of the ACS lines. For this purpose, the \bk--space data for every realization in each data set was subsampled with an ACS region containing $32$ lines. In order to study the effect of the kernel size and the acceleration factor, three combinations of parameters were used: $\{R=3, \text{kernel}=[2,3]\}$, $\{R=4, \text{kernel}=[2,3]\}$ and $\{R=3, \text{kernel}=[4,3]\}$. For each scenario, the first realization of each data set was used to compute the GRAPPA kernel and the coil combination vector, which then were used to reconstruct the final image. 
	\item Third, we considered the case of non-uniform subsampling patterns caused by a Variable Density reconstruction. For this purpose, the first image in each data set was subsampled with different acceleration factors in different \bk--space regions (R=2,3 or 4) and an unaccelerated region at the \bk--space center containing $32$ ACS lines. The ACS were used to obtain the different GRAPPA kernels, from which the coil images were reconstructed and used to estimate the coil combination vector. Then, for each realization, the \bk--space coils were subsampled non-uniformly with the same pattern, reconstructed using GRAPPA with the previously computed kernel and linearly combined as in \cite{Walsh00} with the previously computed combination vector. 
\end{enumerate}

For all the experiments, $g$-factor maps were obtained in three different ways: First, we followed a Monte-Carlo strategy to compute the sample standard deviation for each pixel from all the realizations for both the unaccelerated and the accelerated images. The $g$-factor maps are obtained by simply applying its definition;  Second, we directly computed the $g$-factor maps using the image--space  method described in Eq.~\eqref{eq:$g$-factor_combined}; Third, we directly computed the $g$-factor maps using the proposed \bk--space method as in Eq.~\eqref{eq:$g$-factor_final}. The last two methods require as inputs the \bk--space noise matrices ${\bf \Gamma}$ and ${\bf C}$ in the subsampled images, the GRAPPA kernel and the coil-combination vector. For the simulated dataset, all the parameters were known, whereas for the acquired phantom, these matrices needed to be estimated. Since multiple realizations of the same acquisition were available, we estimated them as the sample covariance matrices obtained across all the realizations. Furthermore, since these matrices are assumed to be stationary across the \bk--space, we computed them for every \bk--space sample and averaged them afterwards. However, in order to minimize errors caused by the residual B$_0$ drift after its removal, only the points where no signal was present were considered for the averaging. For an {\em in-vivo} acquisition, the estimation of these matrices should be done from a single image as in \cite{AjaNoiseBook}. However, the objective of this study is to show that the $g$-factor maps can be computed exactly when these matrices are known.

All image reconstruction and $g$-factor maps estimation were performed using Matlab and run on a standard PC with an Intel\textregistered Core\textsuperscript{TM} i5-4210M @2.6 GHz Processor and 7.5 GB of RAM. In the spirit of reproducible research, we provide a software package including the data sets and code that we used, allowing to reproduce all the results described in this paper. It can be downloaded from https://github.com/Inaki4/kSpaceNoiseEstimationGRAPPA.


\section{Results}

\begin{figure}[tb]
\centering
\includegraphics[width=1\columnwidth]{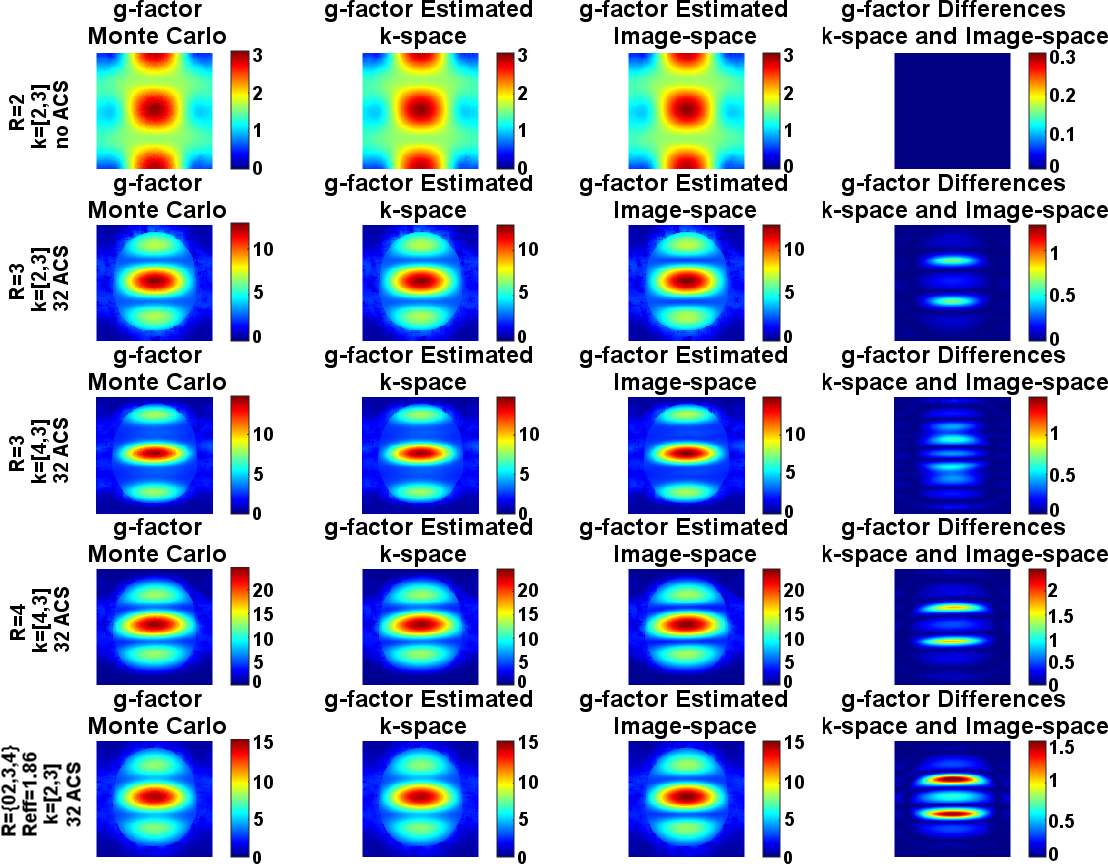}
\caption{$g$-factor maps for the synthetic phantom obtained through a Monte Carlo strategy (first column) and estimated using the proposed \bk--space method (second column) and the image--space method (third column). Last column shows the absolute differences between the exact $g$-factor maps (\bk--space) and the approximated ones (image--space). Several escenarios are studied: uniform undersampling with R=2 and kernel=[2,3] (first row), non-uniform undersampling including 32 ACS with R=3 and kernel=[2,3] (second row), R=3 and kernel=[4,3] (third row), R=4 and kernel=[2,3] (fourth row), and non-uniform undersampling with a Variable Density approach (fifth row).}
\label{fig:SyntheticResults}
\end{figure}
\begin{figure}[tb]
\centering
\includegraphics[width=1\columnwidth]{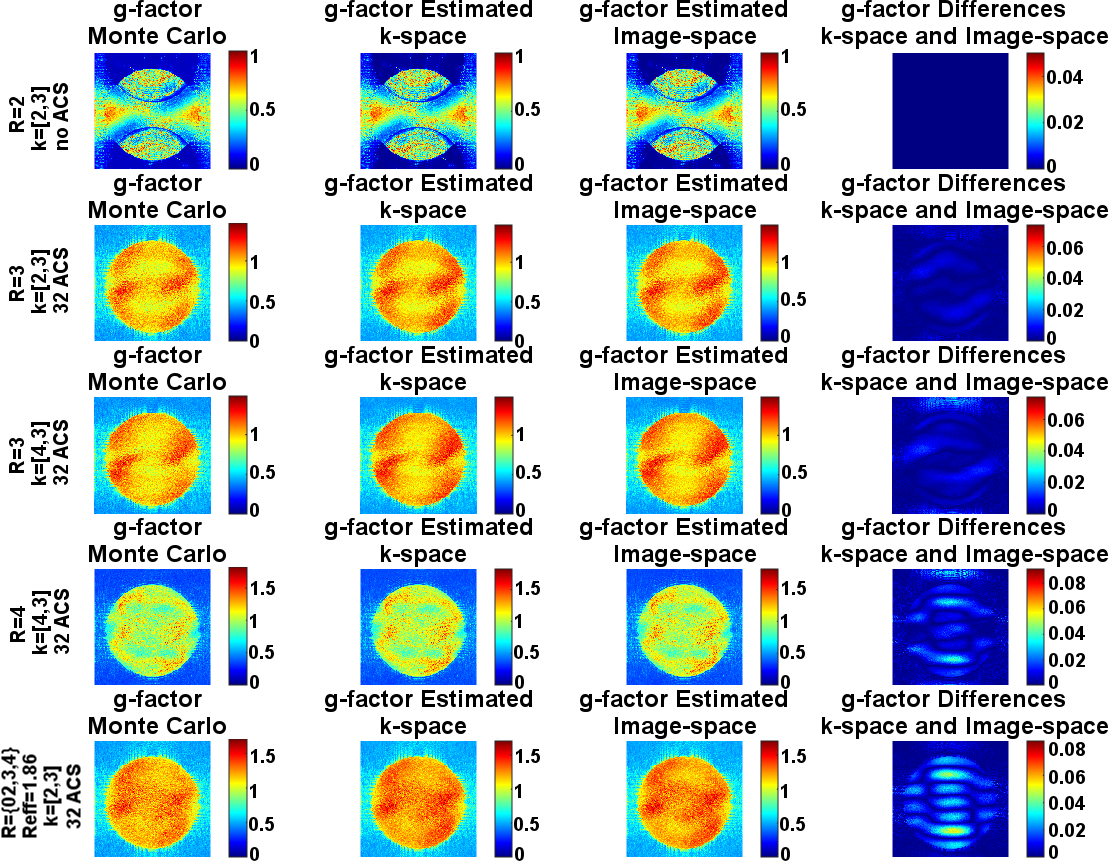}
\caption{$g$-factor maps for the scanned water phantom. Same scenarios and results than for the synthetic phantom are shown.}
\label{fig:RealResults}
\end{figure}

Fig.~\ref{fig:SyntheticResults} shows the $g$-factor for each pixel in the synthetic phantom image reconstructed using GRAPPA with different strategies: uniform undersampling, non-uniform undersampling with ACS and non-uniform undersampling with VD.  
Differences with respect to our method will be considered as approximation errors of the image--space method, due to the analytical nature of our proposal. Consequently, the error introduced by the image--space method is shown in the last column. Fig.~\ref{fig:RealResults} shows the same results for the real phantom.

When the undersampling pattern is uniform across the phase-encoding direction, both methods are completely equivalent, and thus there is no error in the estimation of the $g$-factor maps. However, when the undersampling pattern is not uniform, both methods provide a slightly different estimation as expected. Since the errors introduced by the image--space method arise from the reconstruction of the boundaries between differently-sampled regions, their magnitude will depend on the undersampling pattern. 

As previously reported, the image--space method assumes the \bk--space can be split into independent regions. Importantly, this assumption introduces residual errors due to neglected overlapping between regions, with the number of overlapping lines depending on the size of the kernel. Further, since the acquired lines in the boundaries  that separate every two regions are used to reconstruct both regions, the missing lines in the two regions that are reconstructed from the shared acquired line will be correlated, with the number of correlated lines depending on the acceleration factor in the regions. Figures \ref{fig:SyntheticResults} and \ref{fig:RealResults} show how the error grows when the size of the kernel, the acceleration factor or the number of regions are increased.

Finally, as for computational load, we performed a test with a $256 \times 256$ image using both the image--space and the \bk--space methods. The image--space method required 4.66 seconds, whereas our proposed \bk--space method required 4.06 seconds. Importantly, the Matlab code used in this work has not been optimized for speed, and significant time reduction is likely to be achieved due to the high degree of parallelism in some of the reconstruction steps.


\section{Discussion}

An analytical method for noise characterization in GRAPPA reconstructions has been proposed. The method allows for an exact characterization of noise under the assumptions of stationarity and uncorrelation in the original \bk--space undersampled acquisition. The strength of the method lies in two cornerstones: (1) by operating directly in the \bk--space, we succeed in providing the exact characterization of noise by accounting for all the \bk--space correlations; (2) by exploiting the extensive symmetries and the separability in the reconstruction steps, we overcome the computational challenges related to the very large size of the covariance matrices. As a result, the proposed method provides an exact characterization of noise under the assumptions of stationarity and uncorrelation in the original \bk--space undersampled acquisition.

The accurate modeling of noise provided by the theoretical method is also confirmed by the experiments, were it outperforms previously proposed methods. The Monte-Carlo method proposed in \cite{Robson08} is a versatile approach for noise characterization. However, this method is very time consuming, due to the need to reconstruct multiple images, as well as subject to errors due to the limited number of replicas for the Monte-Carlo estimation. Compared with this approach, our proposal provides an exact characterization of noise and avoids the need to generate and reconstruct multiple replicas of the image.  

The image--space method proposed in \cite{Breuer09,AjaMRM11}, succeeds in providing a close approximation to the actual noise maps, which holds exact for the case of uniform undersampling patterns. However, image--space methods are subject to errors when a non-uniform undersampling pattern is used. By operating directly in the \bk--space, our proposed method is able to avoid these errors and provide an exact estimation in a feasible computation time.

This study has  some limitations: First, the analysis in this paper is focused on GRAPPA reconstruction followed by a linear coil combination. For other coil combination strategies, the final step of the analysis must be redone. If SoS is considered, for instance, the noise distribution can be approximated as a non-Central Chi distribution \cite{AjaMRM11} whose effective values must be calculated. However, this step is a straightforward extension of this manuscript once the final covariance matrices are built; Second, this manuscript focuses on 2D-GRAPPA reconstructions accelerated in one dimension. However, the proposed approach can be easily extended to acquisitions with acceleration in two dimensions, or 3D acquisitions. Third, our analysis is focused on Cartesian acquisitions. However, if non-Cartesian acquisitions (eg: radial or spiral trajectories), are followed by a linear interpolation into a Cartesian grid, our procedure can still be applied, although this process could increase the correlations in \bk--space. Fourth, the proposed method assumes that the kernel used for reconstruction is independent of the noise in the acquired image, which is not exact when the kernel is estimated from the data instead of from a separate pre-scan. However, this assumption is expected to be a good approximation due to the typical overdetermination of GRAPPA weights estimation from the ACS region, and for this reason it has become a common assumption in the literature.

\section{Conclusion}

We have proposed a novel method for noise characterization in GRAPPA reconstructions. Through a careful direct \bk--space analysis that exploits both the symmetry and separability in the reconstruction steps, our method provides an exact noise characterization under the assumptions of stationarity and uncorrelation in the acquired \bk--space.

\section*{Acknowledgments}
The authors acknowledge MICIN for grants TEC2013-44194P and TEC 2014-57428, as well as Junta de Castilla y León for grant VA069U16. The first author acknowledges MINECO for FPI grant BES-2014-069524. 

\bibliographystyle{mrm}
\bibliography{strings,grappa}

\begin{thebibliography}{10}

\bibitem{AjaNoiseBook}
Aja-Fern{\'a}ndez A, Vegas-S{\'a}nchez-Ferrero G.
\newblock Statistical Analysis of Noise in {MRI}.
\newblock Springer, 2016.

\bibitem{Krissian09}
Krissian K, Aja-Fern{\'a}ndez S.
\newblock Noise-{D}riven anisotropic {D}iffusion filtering on {MRI}.
\newblock IEEE Trans Image Process 2011;\hspace{0pt}65:138--145.

\bibitem{Aja13}
Aja-Fern{\'a}ndez S, Brion V, Trist{\'a}n-Vega A.
\newblock Effective noise estimation and filtering from correlated
  multiple-coil {MR} data.
\newblock Magn Reson Imag 2013;\hspace{0pt}31:272--285.

\bibitem{Thunberg07}
Th\"unberg P, Zetterberg P.
\newblock Noise distribution in {SENSE-} and {GRAPPA-} reconstructed images: a
  computer simulation study.
\newblock Magn Reson Imag 2007;\hspace{0pt}25:1089--94.

\bibitem{Saritas11}
Saritas E, Lee J, Nishimura D.
\newblock {SNR-} dependence of optimal parameters for apparent diffusion
  coefficient measurements.
\newblock IEEE Trans Med Imag 2011;\hspace{0pt}2:424--437.

\bibitem{Rabanillo16}
Rabanillo I, Hernando D, Aja-Fern{\'a}ndez S.
\newblock Variation of noise in multi-echo {MRI} acquisitions using {P}arallel
  {I}maging.
\newblock Magn Reson Med 2011;\hspace{0pt}65:138--145.

\bibitem{Ghugre05}
Ghugre N, Enriquez C, Coates T, Nelson M, Wood J.
\newblock Improved r2* measurements in myocardial iron overload.
\newblock J Magn Reson Imaging 2005;\hspace{0pt}23:9--16.

\bibitem{Veraart11}
Veraart J, Poot D, Hecke WV, Blockx I, der Linden AV, Verhoye M, Sijbers J.
\newblock More accurate estimation of diffusion tensor parametersusing
  diffusion kurtosis imaging.
\newblock Proceedings of the 33rd European Society for Magnetic Resonance in
  Medicine and Biology Scientific Meeting (ESMRMB ’16)
  2016;\hspace{0pt}29:400.

\bibitem{Pruessmann99}
Pruessmann KP, Weiger M, Scheidegger MB, Boesiger P.
\newblock {SENSE}: Sensitivity encoding for fast {MRI}.
\newblock Magn Reson Med 1999;\hspace{0pt}42(5):952--62.

\bibitem{Griswold02}
Griswold M, Jacob P, Heidermann R, Nittka M, Jellus V, Wang J, Kiefer B, Haase
  A.
\newblock Generalized autocalibrating partially parallel acquisitions
  ({GRAPPA}).
\newblock Magn Reson Imag 2002;\hspace{0pt}47:1202--10.

\bibitem{Robson08}
Robson P, Grant A, Madhuranthakam A, Lattanzi R, Sodickson D, McKenzie C.
\newblock Comprehensive quantification of signal-to-noise ratio and g-factor
  for image-based and k-space-based parallel imaging reconstructions.
\newblock Magn Reson Med 2008;\hspace{0pt}60:895.

\bibitem{Breuer09}
Breuer F, Kannengiesser S, Blaimer M, Seiberlich N, Jakob PM, Griswold MA.
\newblock General formulation for quantitative {G}-factor calculation in
  {GRAPPA} reconstructions.
\newblock Magn Reson Med 2009;\hspace{0pt}62(3):739--746.

\bibitem{AjaMRM11}
Aja-Fern\'{a}ndez S, Trist\'an-Vega A, Hoge WS.
\newblock Statistical noise analysis in {GRAPPA} using a parametrized
  non-central chi approximation model.
\newblock Magn Reson Med 2011;\hspace{0pt}65(4):1195--1206.

\bibitem{Walsh00}
Walsh D, Gmitro A, Marcelling M.
\newblock Adaptive reconstruction of phased array {MR} imagery.
\newblock Magn Reson Med 2000;\hspace{0pt}43(5):682--690.

\bibitem{Roemer90}
Roemer P, Edelstein W, Hayes C, Souza S, Mueller O.
\newblock The {NMR} phased array.
\newblock Magn Reson Med 1990;\hspace{0pt}16:192--225.

\bibitem{Brau08}
Brau A, Beatty P, Skare S, Bammer R.
\newblock Comparison of reconstructionaccuracy and efficiency among
  autocalibrating data-driven parallel imaging methods.
\newblock Magn Reson Med 2008;\hspace{0pt}59:382--395.

\bibitem{Heidemann07}
Heidemann R, Griswold M, Seiberlich N, Nittka M, Kannengiesser S, Kiefer B,
  Jakob P.
\newblock Fast method of 1{D} non-{C}artesian parallel imaging using {GRAPPA}.
\newblock Magn Reson Med 2007;\hspace{0pt}57(6):1037--46.

\bibitem{Goodman63}
Goodman N.
\newblock Statistical analysis based on a certain multivariate complex gaussian
  distribution ({A}n introduction).
\newblock Annals {M}ath {S}tatist 1963;\hspace{0pt}34:152--177.

\bibitem{DavisBook}
Davis P.
\newblock Circulant matrices.
\newblock Wiley, 1979.

\bibitem{Collins98}
Collins D, Zijdenbos A, Kollokian V, Sled J, Kabani N, Holmes C, Evans A.
\newblock Design and construction of a realistic digital brain phantom.
\newblock IEEE Trans Med Imag 1998;\hspace{0pt}17(3):463--468.

\bibitem{AjaSENSE}
Aja-Fern{\'a}ndez S, Vegas-S{\'a}nchez-Ferrero G, Trist{\'a}n-Vega A.
\newblock Noise estimation in parallel {MRI}: {GRAPPA} and {SENSE}.
\newblock Magn Reson Imag 2014;\hspace{0pt}32(3):281--290.

\end{thebibliography}

\end{document}